\documentstyle[aps,pre,multicol,epsfig]{revtex}
\draft
\input epsf
\begin{document}
\title{\bf A microscopic $2D$ lattice model of dimer granular compaction with
friction}
\author{C. Fusco${}^1$\thanks{Author to whom correspondence should be
addressed. Electronic address: fusco@sci.kun.nl.}, A. Fasolino${}^1$,  
P. Gallo${}^2$, A. Petri${}^{3,4}$ and M. Rovere${}^2$
}
\address{
${}^1$ Department of Theoretical Physics, University of Nijmegen,\\ 
Toernooiveld 1, 6525 ED Nijmegen, The Netherlands\\  
${}^2$ Dipartimento di Fisica, Universit{\`a} Roma
Tre, \\ 
and Istituto Nazionale per la Fisica della Materia,  Unit{\`a} di Ricerca
Roma Tre, \\ Via della Vasca Navale 84, I-00146 Roma, Italy \\
${}^3$ Consiglio Nazionale delle Ricerche,
Istituto di Acustica ``O. M. Corbino'',\\
Area della Ricerca di Roma Tor Vergata,\\
Via del Fosso del Cavaliere 100, 00133 Roma, Italy\\
${}^4$ INFM,  Unit{\`a} di Roma 1, Piazzale Aldo Moro 2, 00185 Roma, Italy
}

\maketitle

\begin{abstract} 

We study by Monte Carlo simulation the compaction dynamics of hard dimers
in $2D$ under the action of gravity, subjected to vertical and horizontal
shaking,  considering also the case in which a friction force acts
for horizontal displacements of the dimers. These forces are modeled by
introducing effective probabilities for all kinds of moves of the particles.
We analyze the dynamics for different values of the time $\tau$ during
which the shaking is applied to the system and for different intensities of
the forces. It turns out that the density evolution in time follows 
a stretched exponential behavior if $\tau$ is not very large, while
a power law tail develops for larger values of $\tau$. Moreover, in the
absence of friction, a critical value $\tau^*$ exists which signals
the crossover between two different regimes: for $\tau < \tau^*$ the
asymptotic density scales with a power law of $\tau$, while for
$\tau > \tau^*$ it reaches logarithmically a maximal saturation value.
Such behavior smears out when a finite friction force is present. In this
situation the dynamics is slower and lower asymptotic densities
are attained. In particular, for significant friction forces,
the final density decreases linearly with the friction coefficient. 
We also compare the frictionless single tap dynamics to the sequential tapping
dynamics, observing in the latter case an inverse logarithmic behavior of the
density evolution, as found in the experiments.

\end{abstract}

\begin{multicols}{2}
\section{Introduction}
\label{sec.intro}

The process of compaction in granular media attracts a great deal of attention
in the scientific community, in particular amongst physicists and chemical 
engineers. 
In fact it displays features that are general enough to make 
it suitable for investigation through models that are  
relatively simple, at least for a non equilibrium process. 
Granular compaction therefore appears appealing for 
testing new promising and unifying ideas in the field of 
disordered systems~\cite{Granular,Proc-Gran,Jaeger1,Jaeger2}.  

Grains poured into a  vessel fill it with loose arrangements and
relatively low densities.
The action of external perturbations, like shaking and tapping, in the
presence of an external driving, like the gravity field, leads to a
very slow increase of density through a rich phenomenology, 
displaying different regimes and both reversible and irreversible dynamical 
phases~\cite{Knight,Nowak1,Nowak2,Ben-Naim,Talbot,Linz,Edwards,Brey,Gavrilov,Philippe,Head1,Head2,Luding,Hertzsch}.
A number of different lattice models have been recently investigated 
in order to unravel the microscopic mechanisms producing
such  behaviors, and to clarify to which extent they may be
considered general, also in relation to other disordered systems like
glasses~\cite{Nicodemi1,Nicodemi2,Nicodemi3,Caglioti,Barrat1,Barrat2,Sellitto,Kob,Arenzon}.    

In this work we introduce a lattice gas model 
for dimer compaction which includes, besides the vertical, also 
horizontal shaking and friction. Friction,  
although playing a fundamental role in real granular
systems, has not been introduced so far in models for compaction.
Horizontal shaking has been considered only experimentally
for non Brownian spheres where it seems to favor local
dense crystalline order~\cite{Pouliquen,Nicolas}.
Highly anisotropic granulars, namely rods, have been recently 
investigated under vertical shaking. In contrast to sphere packings
which tend to end up in disordered configurations, evidence is
found that the particle anisotropy drives ordering~\cite{Villarruel}.
We try to incorporate in our model, albeit in a simplified manner, all these
aspects, namely particle anisotropy, horizontal shaking and frictional 
dynamics, which we find to affect significantly the compaction process.

In particular, our model is based on the diffusional dynamics of dimers,
considered as rigid and non overlapping
particles which occupy two lattice sites on a square lattice. 
The model has no quenched disorder, but for this type of dynamics glass-like 
properties may be expected~\cite{Fusco1},
because of the onset of geometrically frustrated configurations.
Since the number of states with the closest packed density 
is exponentially large in the lattice size \cite{Kasteleyn,Fisher}, this 
model is intrinsically different from the other 
lattice models recently investigated~\cite{Nicodemi1,Nicodemi2,Nicodemi3,Caglioti,Barrat1,Barrat2,Sellitto} and is suitable to understand 
how horizontal shaking and friction may influence the
general characteristics attributed to the compaction process.
  
In the next  section we describe in detail the model and the Monte Carlo
method that we used. In Sec.~\ref{sec.dynamics} we discuss the results
for the compaction dynamics without friction and illustrate the 
evolution of the packing structure.
In Sec.~\ref{sec.friction} we introduce the effect of friction
and show its crucial role in the dynamical behavior.
The final section is devoted to the conclusions.   

\section{Model} 
\label{sec.model}

We consider a square lattice of $N=L\times L$ sites, with lateral periodic 
boundary conditions, an open boundary at the top and a rigid wall at the 
bottom. Elongated particles occupying two consecutive lattice sites (dimers) 
are inserted from the top one at the time with random (horizontal or vertical)
orientation, letting them fall down keeping their orientation fixed, 
until reaching a stable position. 
\begin{figure}
\centering\epsfig{file=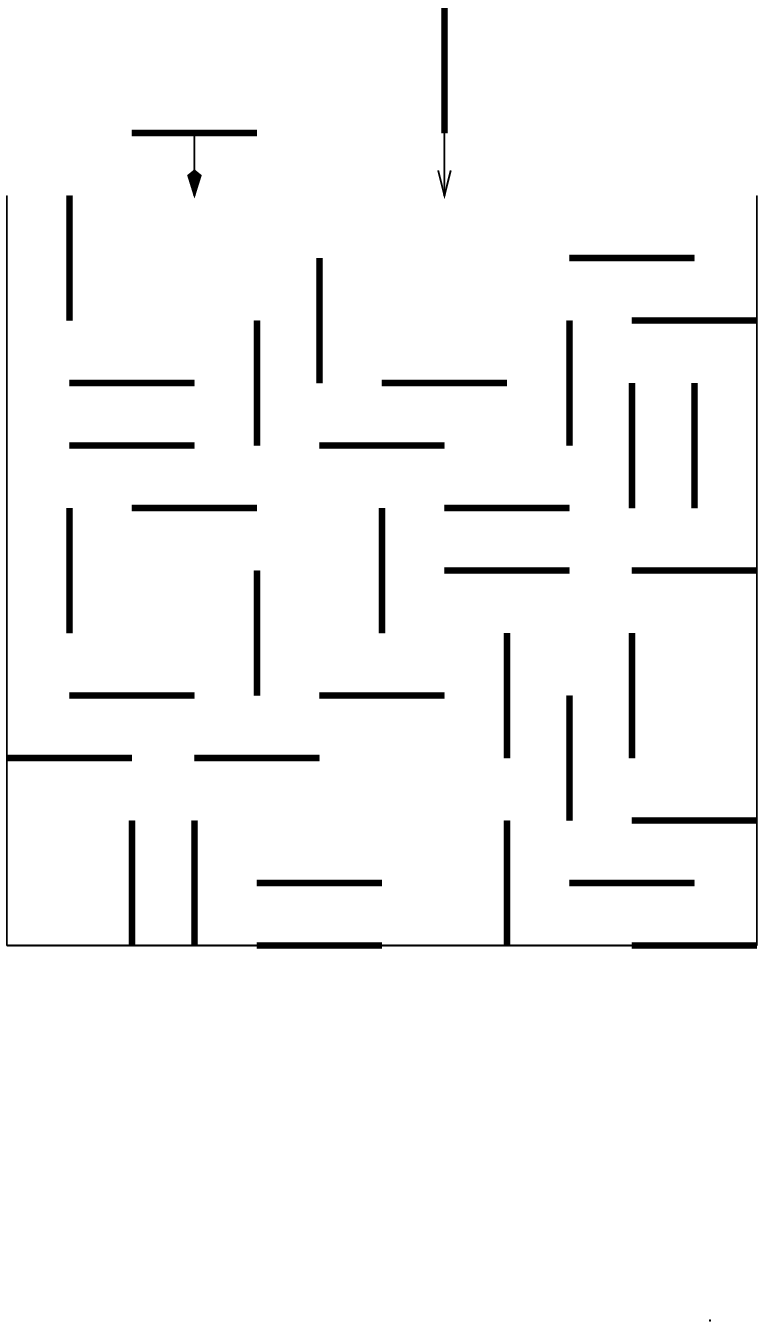, width=7cm, height=10cm}
\caption{Schematic picture of the lattice model studied. The arrangement of 
the particles is the result of the initial sample preparation. Note that 
lateral periodic boundary conditions have been imposed.}
\label{f.1}
\end{figure}
In this way we are able to prepare the system in an initial 
state which is saturated, i.e. no more particles can be put in, and 
with statistically reproducible density $\rho_0$, i.e. characterized by 
a precise mean value, $\rho_0\simeq 0.587$, corresponding to a random loose 
packing.
In the present model, particles are subjected only to geometrical 
interactions, and the non-overlapping condition
produces strong constraints on their relative positions (see Fig.~\ref{f.1}). 
The aim of this paper is to study the system in presence of gravity and 
external vibrations, with the possibility of taking into account also a 
friction-like force for horizontal displacements. 
For this purpose we perform Monte Carlo (MC) simulations introducing a random 
diffusive dynamics in our model, which mimics the aforementioned forces and 
preserves the geometrical constraints. 
At this stage, we consider separately horizontal and vertical dimers and keep 
their orientation fixed without allowing rotations. We plan to introduce this
feature, which would make the model more realistic, in the future. 
We consider in detail a single tap applied to the system 
for a fixed time $\tau$, comparing it with the multiple
tapping for selected cases (see~\cite{Nicodemi1}). 
We first describe the dynamics without friction.
The dynamics can be divided into two stages:
$(i)$ for $t<\tau$ particles can move horizontally (left or right with 
probability $p_h/2$) and vertically (upwards with probability $p_{up}$ or
downwards with probability $p_{down}$);
$(ii)$ for $t>\tau$ only downward movement of particles is possible
(i.e. $p_{up}=p_{h}=0$ and $p_{down}=1$).
We note that $\tau^{-1}$ can be thought of as the quench rate of an initially 
annealed system.
The probabilities for the different moves are normalized: 
$p_{h}+p_{up}+p_{down}=1$. Physically $p_{down}$ corresponds to the action of 
gravity, while $p_{h}$ and $p_{up}$ represent respectively the horizontal and
vertical component of the external tap. The functional form chosen 
for the time dependence of $p_{h}$ and $p_{up}$ is~\cite{Nicodemi1}
\begin{equation}
\label{e.ph}
p_{h}(t)=p_{h}^0(1-t/\tau)\theta(\tau-t),
\end{equation}
where $\theta(x)$ is the Heaviside function
(the same expression holds for $p_{up}(t)$, with $p_{up}^0$ instead of 
$p_{h}^0$).
$p_{down}(t)$ is determined from the normalization condition:
\begin{equation}
p_{down}(t)=1-p_{up}(t)-p_{h}(t)=p_{down}^0+(p_{up}^0+p_{h}^0)t/\tau
\end{equation}
where $p_{down}^0\equiv 1-(p_{up}^0+p_{h}^0)$. Since $p_{down}(t)\ge 0$ 
$\forall t$ it must be $p_{up}^0+p_{down}^0\le 1$. If not stated explicitly
otherwise, we assume $p_{down}^0=p_{up}^0$, so $p_{h}^0$ results to be
$p_h^0=1-2p_{up}^0$, that is only one independent input parameter is needed. 
In each MC move one extracts a particle with uniform probability, chooses
a move for this particle according to the values of $p_{h}$, $p_{up}$ and 
$p_{down}$, and performs the move if all the geometrical constraints are
satisfied. One MC time step (MCS) corresponds to $N$ attempted moves. 
In the following, time is always given in units of MCS.
To save CPU time we used an algorithm in which the attempted moves are always 
successful, and consistently updated time through probabilistic arguments
(for the details of our computation see Ref.~\cite{Fusco2}
on a reaction-diffusion model for dimers).
We performed our MC simulations on a lattice with $L=100$, for which we 
checked that finite size effects are negligible.

As a next step we introduce friction between adjacent 
horizontal layers in the material in the following simplified way: 
the effective probability of making a horizontal move is set to 
$$
p_{h}^{eff}(t)=p_{h}(t)-\mu(p_{down}(t)-p_{up}(t))=
$$
\begin{equation}
=[p_{h}^0-\mu(1-p_{h}^0-2p_{up}^0)-[p_{h}^0+\mu(p_h^0+2p_{up}^0)]t/\tau]
\theta (t_0(\mu)-t)
\end{equation}
where 
\begin{equation}
t_0(\mu)=\tau \left[ \frac {p_{h}^0-\mu(1-p_{h}^0-2p_{up}^0)}
{p_{h}^0+\mu(p_{h}^0+2p_{up}^0)} \right],
\end{equation}
$\mu$ acts as a friction coefficient ($\mu>0$) and the 
``friction force'' is proportional to the load, represented by the net
vertical force, reducing the probability of the move. 
This assumption implies that the frictional force occurs only for horizontal
moves, independently of whether an underlying dimer (i.e. a dimer in the 
adjacent lower row) is present. We are aware that this is a simplified
approximation, but it is meant as a first step in describing this complex 
process. To ensure $p_{h}^{eff}\ge 0$ at $t=0$, $\mu$ has to satisfy
the condition
$$
\mu\le\frac{p_{h}^0}{(1-p_{h}^0-2p_{up}^0)}. 
$$
In addition, for the friction force to give a negative contribution to 
$p_{h}$, one must have $p_{down}(t)\ge p_{up}(t)$ $\forall t$ 
(this is automatically satisfied with our choice $p_{down}^0=p_{up}^0$). 
Note that the normalization of the frictionless probabilities is still 
$p_{up}+p_{down}+p_{h}=1$. In other terms we have introduced a ``sticking'' 
probability (probability that the particle does not move) $p_{s}$, 
given by $p_{s}=\mu(p_{down}-p_{up})$.
If we put $\mu=0$ we recover the frictionless case.

\section{Compaction dynamics without friction}
\label{sec.dynamics}

We have analyzed the time behavior of the mean density of the system $\rho(t)$
(number of occupied sites normalized to the total number of sites)
by measuring it in the lower $30\%$ of the box at time $t$. 
In this way we are sure to measure density at the bulk, since fluctuations can
increase significantly in the proximity of the open boundary, resulting in a
quite complicated compaction behavior in this region~\cite{Barrat1}. 
Since the statistical
fluctuations of the density can be quite relevant we have performed several MC
realizations (up to $500$) of the process in order to obtain 
reliable results. After preparing the system in its low density configuration
($\rho_{0}\equiv\rho(t=0)\simeq 0.587$), the diffusive dynamics starts as 
described in the previous section. $p_{h}^0$ is chosen as input parameter
and consequently $p_{up}^0=p_{down}^0=(1-p_h^0)/2$. 
We stop the simulations when a steady asymptotic value $\rho_{\infty}$ 
of the density is reached.
We have studied the density evolution for different values of $p_{h}^0$ and 
of the  shaking time $\tau$, in order to find out if some kind of 
scaling law, describing the behavior of the final density as a function of 
these parameters, exists.
As far as we know, no systematic study concerning this point has 
been performed.

\begin{figure}
\centering\epsfig{file=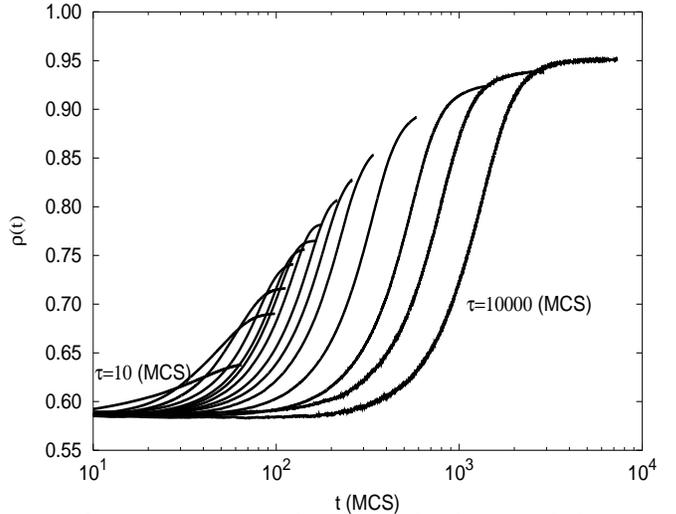, width=9cm, height=7cm}
\caption{Temporal behavior of density in the frictionless case for different 
values of the shaking time $\tau$ (from left to right $\tau=10,30,50,80,103,
125,150,210,280,400,800,2000,4000,10000$). 
The horizontal shaking amplitude is $p_{h}^0=0.1$.}
\label{f.2}
\end{figure}

\begin{figure}
\centering\epsfig{file=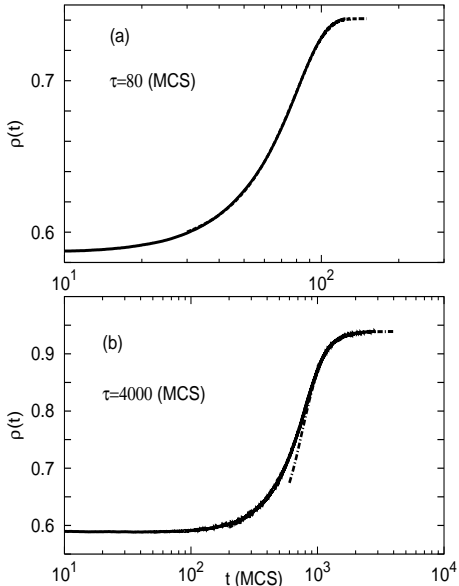, width=10cm, height=8cm}
\caption{Density relaxation in the frictionless case for $\tau=80$ (a) and 
$\tau=4000$ (b). The solid lines are the result of the MC simulation,  
the dashed lines are stretched exponential fits using Eq.~(\ref{e.stretexp})
for $\rho(t)\ge 0.6$, with parameters $C\simeq 1.1$, $t_0\simeq 170$, 
$\tau_0\simeq 250$, $\beta\simeq 9.8$ for $\tau=80$ (a), and $C\simeq 1.1$,
$t_0\simeq 2900$, $\tau_0\simeq 3600$, $\beta\simeq 9.7$ for $\tau=4000$ (b). 
The dot-dashed line in (b) is a fit for $t\ge 1000$ 
according to Eq.~(\ref{e.powlaw}) with parameters $B\simeq 0.018$, 
$\tau\simeq 330$ and $\alpha\simeq 4.95$. 
The horizontal shaking amplitude is $p_{h}^0=0.1$.}
\label{f.3}
\end{figure}

The time behavior of the density for several choices of $\tau$ is illustrated 
in Fig.~\ref{f.2}, where a relatively low value of $p_{h}^0$ has been used
($p_{h}^0=0.1$). 
We have observed that the dynamics gets drastically slower 
when $\tau$ increases; actually, because of our shaking rule, 
Eq.~(\ref{e.ph}), when a long tap is applied 
the density does not significantly change in the first steps of 
the process, but the evolution takes place on a much wider time scale and 
finally a larger value of the asymptotic density is achieved. 
We have tried to fit different functional forms to the data in 
Fig.~\ref{f.2}, looking in particular at the relaxation functions proposed 
in the seminal experimental paper of Ref.~\cite{Knight}. 
It is clear that the observed dynamical behavior 
is very complex and is not compatible with a single relaxation time, i.e. a 
simple exponential. We have found that the commonly claimed 
inverse logarithmic relaxation does not hold for our system in the single tap
case. Instead, the most suitable functional form for our data is a stretched 
exponential, as proposed by Nicodemi {\em et al.}~\cite{Nicodemi1} for the 
single tapping dynamics:
\begin{equation}
\label{e.stretexp}
\rho(t;\tau,p_{h}^0)=\rho_{\infty}(\tau,p_{h}^0)-C(\rho_{\infty}(\tau,p_h^0)-
\rho_{0})\exp[-((t+t_0)/\tau_0)^\beta]
\end{equation}   
where the fit parameters $C$, $t_0$, $\tau_0$ and $\beta$ depend in 
principle both on $\tau$ and $p_{h}^0$. Actually, it turns from our fits that 
$C\simeq 1.1$ and $\beta\simeq 10$ are almost independent of $\tau$ and 
$p_{h}^0$.  
In Fig.~\ref{f.3}(a) we show the density evolution 
for a small value of $\tau$. 
As it can be seen, the intermediate-long
time behavior can be accurately fitted by Eq.~(\ref{e.stretexp}) (all
fits have been performed for $\rho\ge 0.6$). As $\tau$ increases, however, 
Eq.~(\ref{e.stretexp}) fails to reproduce the long time regime. In fact, a
power law tail develops, indicating that the compaction process slows down at
long times for large shaking duration. The fitting function that we have used
for the long time behavior is
\begin{equation}
\label{e.powlaw}
\rho(t;\tau,p_{h}^0)=\rho_{\infty}(\tau,p_{h}^0)-
\frac{\rho_{\infty}(\tau,p_{h}^0)-\rho_0}{1+B(t/\tau_0^{\prime})^{\alpha}}
\end{equation}
This is illustrated in Fig.~\ref{f.3}(b) for $\tau=4000$. Fit 
(\ref{e.powlaw}) has been applied for $\tau>1000$. The exponent $\alpha$ of 
this power law can be considered independent of $\tau$ ($\alpha\simeq 5$).
However the stretched exponential function still describes the 
density behavior in the intermediate time regime well.

\begin{figure}
\centering\epsfig{file=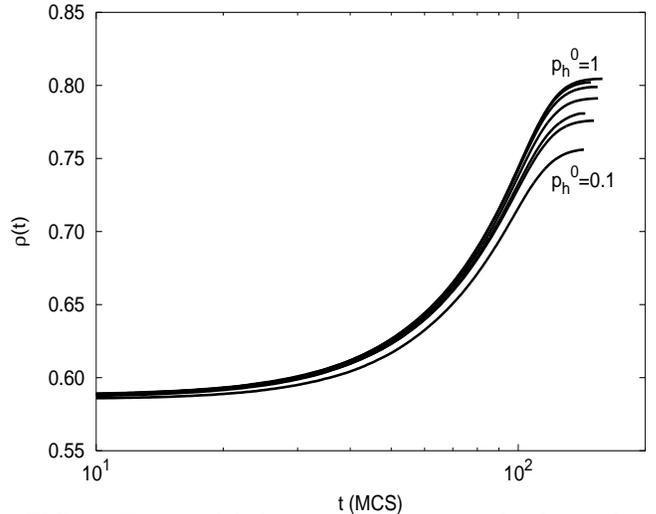, width=9cm, height=7cm}
\caption{Temporal behavior of density in the frictionless case for different 
values of the horizontal shaking amplitude $p_{h}^0$ (from bottom to top 
$p_{h}^0=0.1,0.2,0.3,0.4,0.6,0.8,1$). The shaking time is $\tau=103$.}
\label{f.4}
\end{figure}

We have also studied the density relaxation for different values of 
$p_{h}^0$ at fixed $\tau$ ($\tau=103$). The corresponding plots are shown in 
Fig.~\ref{f.4}, where it is clear that the role of the horizontal shaking
amplitude is crucial in determining the asymptotic density.

\begin{figure}
\centering\epsfig{file=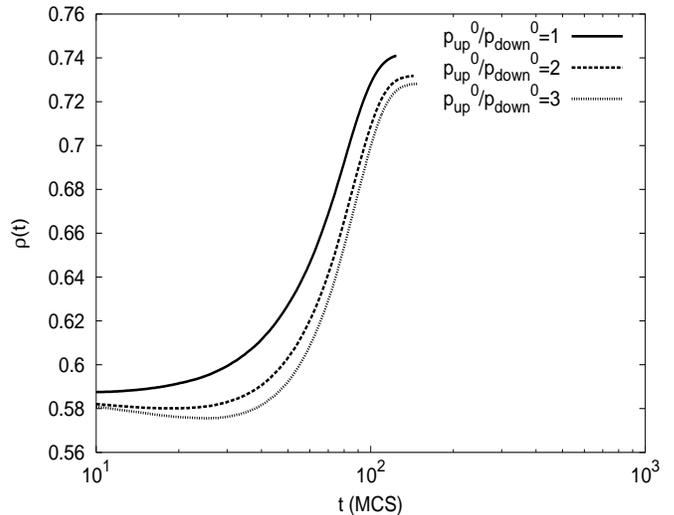, width=9cm, height=7cm}
\caption{Temporal behavior of density in the frictionless case for different
values of the ratio $p_{up}^0/p_{down}^0$: $p_{up}^0/p_{down}^0=1$ (solid 
line), $p_{up}^0/p_{down}^0=2$ (dashed line) and $p_{up}^0/p_{down}^0=3$ 
(dotted line). The shaking time is $\tau=80$ and 
the horizontal shaking amplitude is $p_{h}^0=0.1$.}
\label{f.5}
\end{figure}

Moreover we have considered the effect of different ratios
$p_{up}^0/p_{down}^0$ on the density evolution, which we show in 
Fig.~\ref{f.5}. When $p_{up}^0>p_{down}^0$ we observe a decompaction at the 
beginning of the process, which becomes more pronounced if the vertical tap 
is stronger. 
This reflects the fact that, in the initial transient, 
density decreases at the bottom of the container for a strong tap. 
Such a decompaction upon vertical acceleration increase seems to be a genuine
feature of two-dimensional systems, as claimed in~\cite{Duran}. 
The saturation density is also sensitive
to variations of $p_{up}^0$ and decreases for increasing 
$p_{up}^0/p_{down}^0$. However, the scaling behavior does not change and 
the density evolution can still be described by Eq.~\ref{e.stretexp}, but 
with a smaller value of $\beta$ with respect to the case $p_{up}^0=p_{down}^0$
($\beta\simeq 9.8$ for $p_{up}^0/p_{down}^0=1$, $\beta\simeq 7.6$ for 
$p_{up}^0/p_{down}^0=2$ and $\beta\simeq 6.5$ for $p_{up}^0/p_{down}^0=3$).
 
\begin{figure}
\centering\epsfig{file=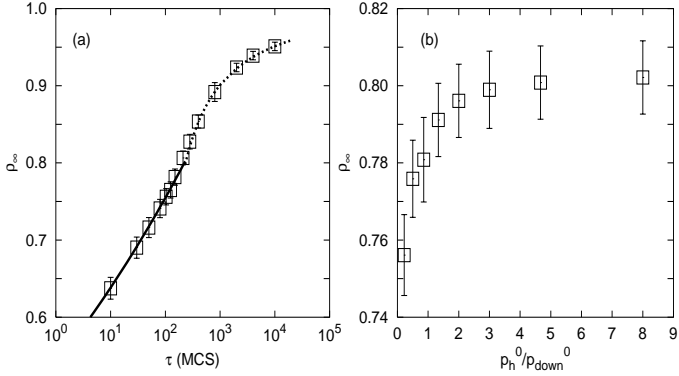, width=9cm, height=7cm}
\caption{Saturation density $\rho_{\infty}$ in the frictionless case 
as a function of the shaking time $\tau$ (a) and the ratio 
$p_{h}^0/p_{down}^0$ (b). 
The squares are the numerical data, while the solid line and dotted line in
(a) are respectively the fits for $\tau<\tau^*$ and  $\tau>\tau^*$ given
in Eq.~\ref{e.fitrhoinfty}. The horizontal shaking amplitude in (a) is 
$p_{h}^0=0.1$ and the shaking time in (b) is $\tau=103$.}
\label{f.6}
\end{figure}

In order to gain a better understanding of the degree of 
compaction of the system, we investigated the dependence of the saturation
density $\rho_{\infty}$ on $\tau$ and $p_{h}^0$. The data for $\rho_{\infty}$
vs. $\tau$ are plotted in Fig.~\ref{f.6}(a). 
A crossover from a power law 
scaling to a logarithmic behavior is observed at a characteristic 
value $\tau^*\simeq 200$:
$$
\rho_{\infty}(\tau)=(\tau/\tau_{l})^{\delta} \qquad {\rm for}\  \tau<\tau^*
$$
\begin{equation}
\label{e.fitrhoinfty}
\rho_{\infty}(\tau)=\rho_{m}-B/\ln(\tau/\tau_{r}) \qquad {\rm for}\  
\tau>\tau^*
\end{equation}  
where $\tau_l\simeq 4.5\cdot 10^3$, $\delta\simeq 0.073$, $\rho_m\simeq 1$, 
$B\simeq 0.6$ and $\tau_r\simeq 18$.
This means that for large values of $\tau$ the process gets slower and slower,
as observed before, eventually reaching a final value of the density
$\rho_{m}\simeq 1$ for $\tau\rightarrow\infty$ 
(which corresponds to about the maximal density for the 
system~\cite{Fusco2}). The variation of $\rho_{\infty}$ with respect to the 
ratio $p_h^0/p_{down}^0$ is illustrated in Fig.~\ref{f.6}(b). 
The increase of $\rho_{\infty}$ is more pronounced for low values of $p_h^0$,
while it tends to saturate afterwards. This compares qualitatively well 
to the experimental results for a horizontally shaken box filled 
with beads~\cite{Pouliquen}, where it was found that for low filling rates
the packing crystallizes upon increase of the adimensional parameter 
$\Gamma=A\omega^2/g$
(where $A$ and $\omega$ are the amplitude and pulsation of the vibration and
$g$ is the gravitational acceleration), which roughly corresponds to  
$p_{h}^0/p_{down}^0$.

\begin{figure}
\centering\epsfig{file=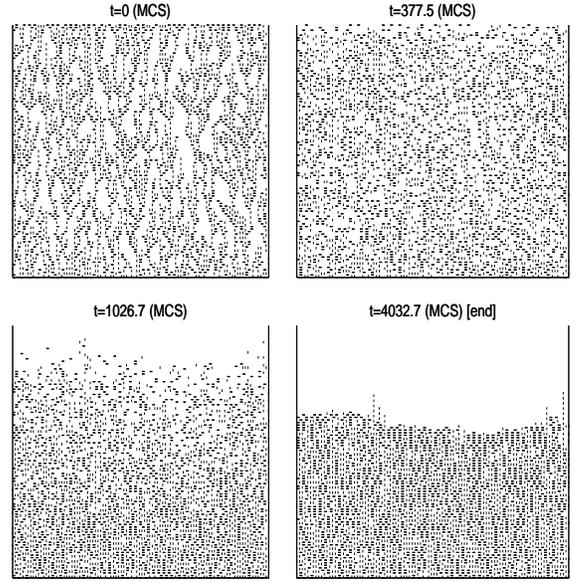, width=9cm, height=8cm}
\caption{Snapshots of one particular realization of the packing at different 
times for $\tau=4000$. The horizontal shaking amplitude is $p_{h}^0=0.1$.}
\label{f.7}
\end{figure}

\begin{figure}
\centering\epsfig{file=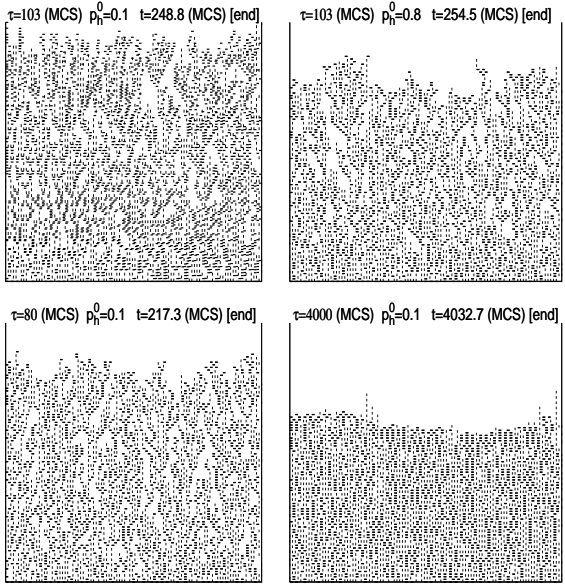, width=9cm, height=8cm}
\caption{Snapshots of the final configuration of the packing for 
two different $p_{h}^0$ and fixed $\tau$ (upper part) and two different 
$\tau$ and fixed $p_{h}^0$ (lower part).} 
\label{f.8}
\end{figure}

In Fig.~\ref{f.7} we show some snapshots of one particular realization of
the compaction process for $\tau=4000$. 
We see how the packing evolves from 
a highly disordered state at $t=0$ to a polycrystalline state, made of
many crystalline ordered domains, at the end of the compaction. 
As time increases the order proceeds 
from the bottom to the top of the packing while empty spaces 
reduce in size. The final state is therefore
characterized by the presence of large 
clusters of horizontal and vertical dimers. In order to demonstrate that 
the shaking amplitude $p_{h}^0$ and  time $\tau$ play a significant role,
we have reported in Fig.~\ref{f.8} a comparison between the final 
configurations for $p_{h}^0=0.1$ and $p_{h}^0=0.8$ at fixed $\tau=103$ 
(upper part), and for $\tau=80$ and $\tau=4000$ at fixed $p_{h}^0=0.1$
(lower part).
We notice that an intense or/and long tap is effective in locally removing 
frozen configurations through collective rearrangements of particles, thus
letting the defects migrate towards the free surface of the packing.

\begin{figure}
\centering\epsfig{file=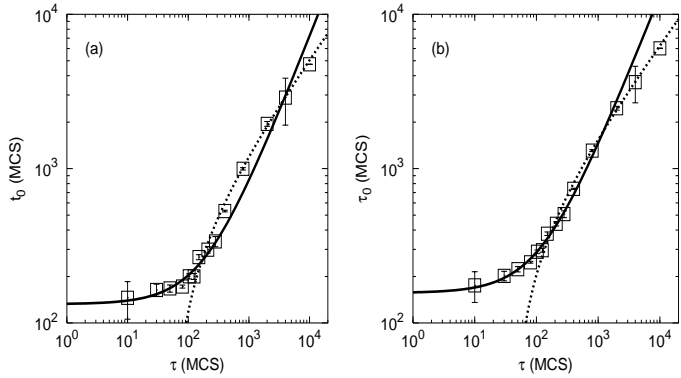, width=9cm, height=7cm}
\caption{Parameters $t_0$ (a) and $\tau_0$ (b) of fit Eq.~(\ref{e.stretexp})
in the frictionless case. The squares are the numerical data, while the 
solid line ($\tau<\tau^*$) and dotted line ($\tau>\tau^*$) in (a) are the 
functions given in Eq.~(\ref{e.fitt0}) and those in (b) are the functions 
given by Eq.~(\ref{e.fittau0}). 
The horizontal shaking amplitude is $p_{h}^0=0.1$.}
\label{f.9}
\end{figure}

We have also investigated in detail the behavior of the parameters of the 
stretched exponential fit (Eq.~(\ref{e.stretexp})) $t_0$ and $\tau_0$, 
which is reported in Fig.~\ref{f.9}. 
We have observed a change of behavior
in correspondence of the critical value of the shaking time $\tau^*$,
the same value determining the change in the behavior 
of $\rho_{\infty}$ vs. $\tau$ described in Eq. (\ref{e.fitrhoinfty}).
A similar transition was also signaled by Nicodemi 
{\em et al.}~\cite{Nicodemi1}.
The parameters of the stretched exponential relaxation 
Eq.~(\ref{e.stretexp}) change according to: 
$$
t_0(\tau)=\tau/t_1+t_< \qquad {\rm for}\  \tau<\tau^*
$$
\begin{equation}
\label{e.fitt0}
t_0(\tau)=(\tau/t_2)^{\gamma}+a \qquad {\rm for}\  \tau>\tau^*
\end{equation}          
$$
\tau_0(\tau)=\tau/\tau_1+\tau_< \qquad {\rm for}\  \tau<\tau^*
$$
\begin{equation}
\label{e.fittau0}
\tau_0(\tau)=(\tau/\tau_2)^{\gamma}+b \qquad {\rm for}\  \tau>\tau^*
\end{equation}
where $t_1\simeq 1.40$, $t_<\simeq 130$, $t_2\simeq 7\cdot 10^{-5}$, 
$\gamma\simeq 0.47$, $a\simeq -700$, 
$\tau_1\simeq 0.76$, $\tau_<\simeq 150$, $\tau_2\simeq 10^{-4}$, 
$b\simeq -700$. It is remarkable that
both $t_0$ and $\tau_0$ vary linearly with $\tau$ for $\tau<\tau^*$ and both
follow a power law with the same exponent $\gamma$ for $\tau>\tau^*$. This 
could mean that they are both determined by the same relaxation mechanism as 
$\tau$ changes (i.e. if we define $\bar{t}_0=t_0-t_<$, 
$\bar{\tau}_0=\tau_0-\tau_<$, $\tilde{t}_0=t_0-a$ and 
$\tilde{\tau}_0=\tau_0-b$, we see from Eqs.~(\ref{e.fitt0}) and
(\ref{e.fittau0}) that $\bar{t}_0/\bar{\tau}_0$ and 
$\tilde{t}_0/\tilde{\tau}_0$ are constants independent of $\tau$). 

\begin{figure}
\centering\epsfig{file=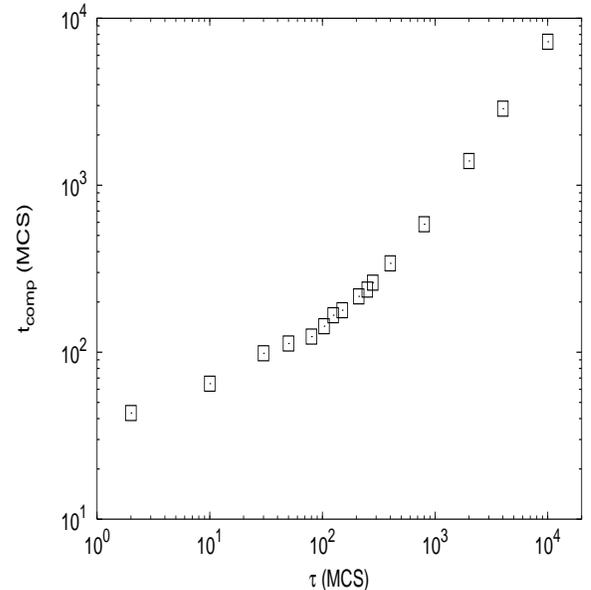, width=8cm, height=8cm}
\caption{Compaction time $t_{comp}$ vs. shaking time $\tau$ in the 
frictionless case. The horizontal shaking amplitude is $p_{h}^0=0.1$.}
\label{f.10}
\end{figure}

We note that the critical shaking time $\tau^*$ physically signals the 
crossover between two different regimes. In fact, a careful inspection of
Fig.~\ref{f.2} shows that, for $\tau<\tau^*$, compaction occurs after the
shaking (when $p_{h}=p_{up}=0$ and $p_{down}=1$) and is mainly driven 
by gravity. 
On the other hand, when $\tau>\tau^*$, the compaction time ($t_{comp}$), 
i.e. the time needed to reach the saturation density, is shorter than $\tau$, 
and vibrations act on the system till the end of the process 
(see Fig.~\ref{f.10}). 
We believe this is responsible for 
the power law tail for large values of $\tau$,
since the combined action of shaking and gravity produces a slowing down of
the dynamics, enabling at the same time to obtain denser packings.
However, we are not sure whether this picture corresponds to a real dynamical
phase transition in the system, and moreover the identification of an
order parameter would be quite problematic. 
Therefore a deeper investigation on 
this point is needed and we hope to stimulate some future experimental and
theoretical works.

\begin{figure}
\centering\epsfig{file=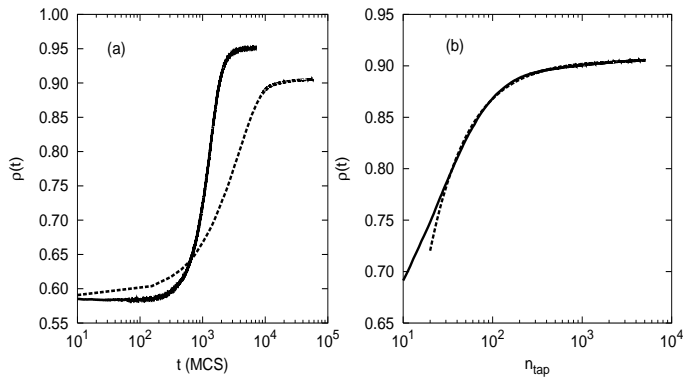, width=9cm, height=7cm}
\caption{Comparison between single and multiple tapping. In (a) the behavior
of density as a function of time is plotted: the solid line
is obtained by a single tap with $\tau=10000$, while the dashed line is 
the result of 5000 taps, each of which applied for a time $\tau/5000$.
In (b) the density is shown as a function of the number of taps ($n_{tap}$): 
the solid line is the result of the simulation and the dashed line is a fit 
according to Eq.~\ref{e.logfit} for $n_{tap}>40$, with parameters 
$c\simeq 47$, $d\simeq 1.2\cdot 10^5$ and $T\simeq 3\cdot 10^4$. 
The horizontal shaking amplitude is $p_{h}^0=0.1$.}
\label{f.11}
\end{figure}

Now we briefly mention the main changes occurring when a sequential tapping 
is applied to the system. Although the single tap dynamics can give 
insight on the dynamical evolution and the relaxation of the system, 
the sequential tapping procedure can be closely related to experimental
situations. In Fig.~\ref{f.11}(a) we compare the density evolution obtained
by a single tap of duration $\tau=10000$ with that obtained by a sequential
tapping applied for the same amount of time $\tau$. In particular we have 
generated $5000$ taps, each of which was applied for a time $\tau/5000$.
From the comparison we deduce that the multiple tapping dynamics is slower, 
suggesting that the fitting form Eq.~\ref{e.stretexp} cannot describe properly
the density evolution in this case. Besides, the saturation density is also
lower. In Fig.~\ref{f.11}(b) we characterize more quantitatively the dynamical
behavior by plotting the density as a function of the number of taps. It turns
out that in this case the intermediate-long time evolution of the density can
be described by the claimed inverse logarithmic law, which was also found in 
the experiments~\cite{Knight}:
\begin{equation}
\label{e.logfit}
\rho(t)=\rho_{\infty}-c\frac{\rho_{\infty}-\rho_0}{1+d\ln(1+t/T)}.
\end{equation}
In this way we can connect directly our model to the experimental data and we 
can also examine the effect of different dynamical shaking procedures on the 
compaction process. In this respect, our results are in agreement with those
of Nicodemi {\em et al.}~\cite{Nicodemi1}, who find a similar change of 
behavior when considering single and multiple tapping. This should clarify to
a certain extent the mechanisms of density relaxation in different dynamical 
situations.

\section{Compaction dynamics with friction}
\label{sec.friction}

\begin{figure}
\centering\epsfig{file=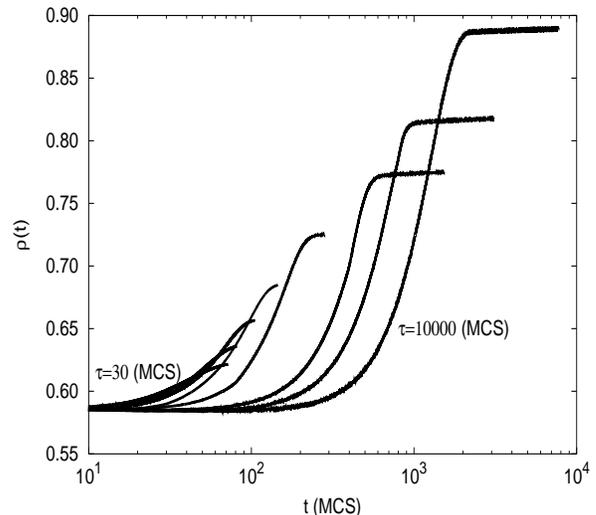, width=8cm, height=7cm}
\caption{Temporal behavior of density for friction coefficient $\mu=0.4$ and
different values of the shaking time $\tau$ (from left to right $\tau=30,
50,80,150,400,2000,4000,10000$).
The horizontal shaking amplitude is $p_{h}^0=0.1$.}
\label{f.12}
\end{figure}

\begin{figure}
\centering\epsfig{file=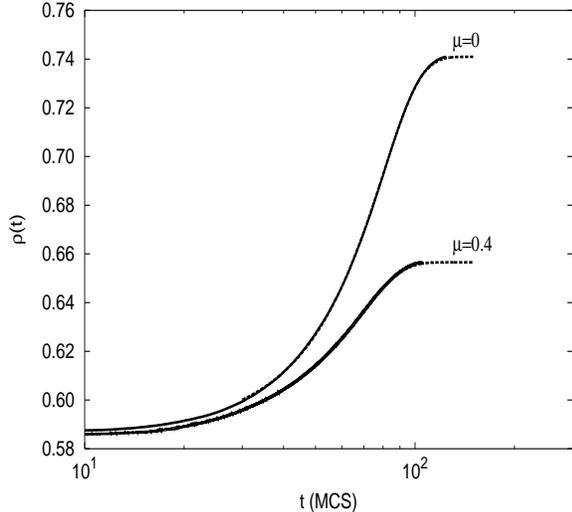, width=8cm, height=7cm}
\caption{Density relaxation for $\mu=0$ (top solid line) and $\mu=0.4$ 
(bottom solid line). The dashed lines are stretched exponential fits using 
Eq.~(\ref{e.stretexp}) for $\rho(t)\ge 0.6$, with parameters $C\simeq 1.1$,
$t_0\simeq 275$, $\tau_0\simeq 400$, $\beta\simeq 9$ for $\mu=0.4$ 
(the parameters for $\mu=0$ can be found in the caption of Fig.~\ref{f.3}).
We note that the major differences between the two curves are due to the 
different values of $\beta$ and $\rho_{\infty}$ 
($\rho_{\infty}(\mu=0)\simeq 0.782$ and $\rho_{\infty}(\mu=0.4)\simeq 0.685$).
The horizontal shaking amplitude is $p_{h}^0=0.1$ and the shaking time is 
$\tau=80$.}
\label{f.13}
\end{figure}

When a frictional force is introduced, as described in Sec.~\ref{sec.intro},
some qualitative and quantitative changes are found. In Fig.~\ref{f.12} the 
time evolution of the density for a fixed value of the friction coefficient 
$\mu$ and several choices of $\tau$ is displayed. 
A comparison with 
Fig.~\ref{f.2} shows that for a same value of $\tau$ a much smaller asymptotic
density is reached, and that a slowing down of the dynamical behavior occurs.
As a consequence, larger values of $\tau$ are needed to obtain the same final
densities as for $\mu=0$. In Fig.~\ref{f.13} a comparison of $\rho(t)$
between $\mu=0$ and $\mu=0.4$ for $\tau=80$ is illustrated. 
It seems that the
same functional form Eq.~(\ref{e.stretexp}) adopted for $\mu=0$ can be used 
to describe the dynamical evolution of the system. In spite of that, 
for $\mu\ne 0$ some slight discrepancies in the tails can be detected 
for small values of $\tau$, not visible at the level of detail of the figure.
This means that the reduced probability for 
horizontal moves affects the compaction mainly in the late stages of the 
process.

We have also examined the $\mu$ dependence of $\rho(t)$. 
The curve for $\mu=0$ is far above the others, and 
furthermore the difference between the curves for low values of $\mu$ is 
more marked. Such differences show up only after an initial transient in 
which the various curves are practically superimposed one on the others
(the curves do not start diverging significantly until the density reaches
about $0.6$). This is in compliance with what we have just said about the 
effectiveness of friction in the asymptotic dynamics.  

\begin{figure}
\centering\epsfig{file=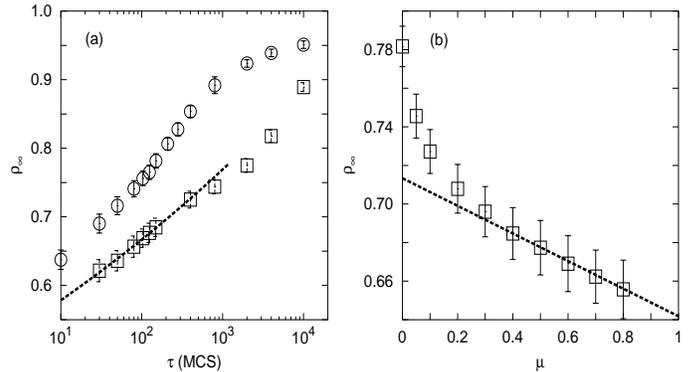, width=9cm, height=7cm}
\caption{Saturation density $\rho_{\infty}$ as a function of (a) $\tau$ for 
$\mu=0.4$ and (b) of $\mu$ for $\tau=150$. The circles in (a)  
and the squares in (a) and (b) are the numerical data for $\mu=0$ and 
for $\mu=0.4$ respectively. The 
dashed lines in (a) and (b) are fits respectively according to 
Eq.~(\ref{e.fitrhoinftyfric}) and Eq.~(\ref{e.fitrhoinftymu}).
The horizontal shaking amplitude is $p_{h}^0=0.1$.}
\label{f.14}
\end{figure}

The behavior of the asymptotic density is depicted in Fig.~\ref{f.14} and 
reveals more interesting aspects. 
In Fig.~\ref{f.14}(a) we have plotted 
$\rho_{\infty}(\tau)$ for $\mu=0$ and $\mu=0.4$. We were able to identify a
power law regime for small $\tau$ also for $\mu=0.4$:
\begin{equation}
\label{e.fitrhoinftyfric}
\rho_{\infty}(\tau)=(\tau/\tau_l^{(\mu)})^{\delta^{(\mu)}}
\end{equation}
with $\delta^{(\mu)}\simeq 0.062$ (not very different from $\delta$ in 
Eq.~(\ref{e.fitrhoinfty})) and $\tau_{l}^{(\mu)}\simeq 6.9 \cdot 10^4$. Thus,
saturation density scales in the same way as for $\mu=0$, with nearly the same
exponent but with a very different relaxation time (more than one order of
magnitude larger), 
i.e. the process is much more sluggish. We did not manage to 
find the functional behavior for larger $\tau$ and it is not very clear 
whether it is possible to define a crossover shaking time, at least 
for the values
of $\tau$ we have considered. For $\tau>1000$ $\rho_{\infty}$ increases more
steeply than for $\mu=0$ and it is likely that one could find another regime
for larger $\tau$. However computational limitations prevented us to explore
in detail the region for $\tau>10000$ (since simulations become more and more
time consuming as $\tau$ increases), and we will address this problem in the 
future. 

\begin{figure}
\centering\epsfig{file=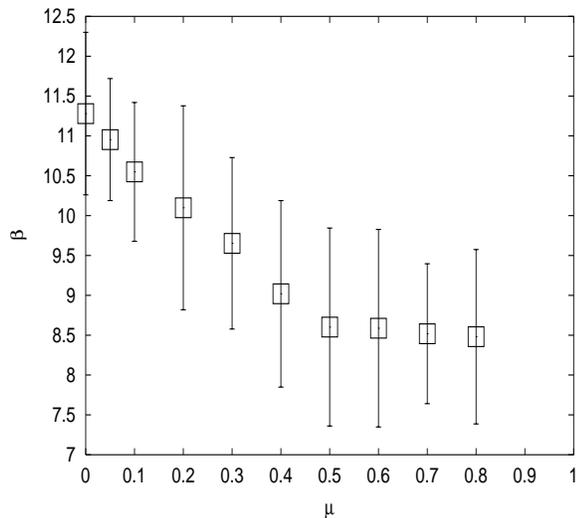, width=8cm, height=7cm}
\caption{Stretched exponent $\beta$ of Eq.~(\ref{e.stretexp}) vs. friction
coefficient $\mu$ for $p_{h}^0=0.1$ and $\tau=150$.}
\label{f.15}
\end{figure}

Finally in Fig.~\ref{f.14}(b) we have analyzed the behavior of 
$\rho_{\infty}$ vs. $\mu$. Interestingly $\rho_{\infty}$ decreases linearly
with $\mu$ for $\mu>0.4$:
\begin{equation}
\label{e.fitrhoinftymu}
\rho_{\infty}(\mu)=A-B\mu, 
\end{equation}
where $A\simeq 0.71$ and $B\simeq 0.073$.
Instead the decrease is more drastic for lower values of $\mu$, as noted 
above. This fact might indicate a change of behavior in the dynamics of the
system as a function of the friction coefficient. For this purpose we show
in Fig.~\ref{f.15} the stretched exponent $\beta$ of Eq.~(\ref{e.stretexp})
vs. $\mu$. 
It rapidly decreases for $\mu<0.4$, but it is almost constant
for larger $\mu$. In other terms, the independence of $\mu$ of the dynamical
evolution reflects itself just in a linear scaling of the final density with
$\mu$.  We hope to elucidate this point more thoroughly in a future work.

\section{Conclusions and perspectives}
\label{sec.conclusions}

We have proposed and discussed a simplified model for dimer compaction, with
the intent to take into account relevant dynamical features, such as 
horizontal shaking and friction forces between particles, analyzing in detail 
their effects on the density time evolution.
The dynamical behavior of our model is very complex with
interesting and novel features.
In the absence of friction, for a single tap, the compaction dynamics cannot 
be interpreted as an inverse logarithmic relaxation but in terms of a 
stretched exponential law, which is a peculiar characteristic of a continuous 
range of time constants~\cite{Knight}.
Thus, the fact that we do not find a logarithmic law but a
stretched exponential relaxation could be a peculiar feature related to the 
single tap dynamics, since Nicodemi {\em et al.}~\cite{Nicodemi1} 
observe the same kind of behavior in a different model with a similar 
dynamics. Furthermore, we are not aware of experiments in which a single tap 
is applied as external perturbation, since they all refer to a tapping 
sequence. Actually, it turns out that the logarithmic behavior found in 
experiments is recovered also in our model when a sequential tapping is 
applied. We therefore might infer that different shaking procedures give rise
to intrinsically different compaction behaviors, driven by different dynamical
mechanisms. However, it still needs to be clarified whether the 
logarithmic law is just a good fit to the experimental data or it has a 
deeper meaning (see for example the discussion in Ref.~\cite{Brey}).

We also find a crossover in the behavior of the asymptotic density
as a function of the shaking time with a slowing down above a critical
shaking time value, indicating a possible
dynamical transition process. 
More detailed studies of this point are planned
for the future in order to reach a better understanding of the
physical mechanism underlying this peculiar behavior. 

Similar to what is found in experiments, horizontal shaking favors 
locally ordered configurations and leads to higher 
compaction~\cite{Pouliquen,Nicolas}. We believe it would be important to 
address further this issue by improving our model, allowing rotations of the 
dimers during the dynamics. 

The results obtained by
introducing the frictional effect in our model indicate the important role of
friction on the compaction dynamics. In particular we find a slowing down
of the dynamical behavior which is more evident in the asymptotic regime.
The relaxation of the density is still described by 
a stretched exponential in the presence of friction, with a  
stretched exponent that becomes constant at large
values of the friction coefficient. 
It would be worthwhile to explore this issue in future works with an improved
modeling of the frictional force.

\end{multicols}


\begin{thebibliography}{99}

\bibitem{Granular}
  {\it Granular Matter}, 
  A. Metha ed. (Springer-Verlag 1994).
 
\bibitem{Proc-Gran}
  Proceedings of the NATO Advanced Study Institute
  on {\it Physics of Dry Granular Media},   
  Eds. H.J. Herrmann et al., Kluwer Academic Publishers,
  The Netherlands (1998).

\bibitem{Jaeger1}
  H.M. Jaeger, S.R. Nagel and R.P. Behringer,   
  Rev. Mod. Phys. {\bf 68}, 1259 (1996).

\bibitem{Jaeger2}
  H.M. Jaeger and S.R. Nagel,   
  Science {\bf 255}, 1523 (1992).


\bibitem{Knight}
  J.B. Knight, C.G. Fandrich, C.N. Lau, H.M. Jaeger and S.R. Nagel,  
  Phys. Rev. E {\bf 51}, 3957 (1995).

\bibitem{Nowak1}
  E.R. Nowak,  J.B. Knight, E. Ben-Naim, H.M. Jaeger and S.R. Nagel,
  Phys. Rev. E {\bf 57}, 1971 (1998).

\bibitem{Nowak2}
  E.R. Nowak,  J.B. Knight, M.L. Povinelli, H.M. Jaeger and S.R. Nagel,
  Powder Technol. {\bf 94}, 79 (1997).

\bibitem{Ben-Naim}
  E. Ben-Naim, J.B. Knight, E.R. Nowak, H.M. Jaeger and S.R. Nagel,
  Physica D {\bf 123}, 380 (1998).

\bibitem{Talbot}
  J. Talbot, G. Tarjus and P. Viot,
  Phys. Rev. E {\bf 61}, 5429 (2000).

\bibitem{Linz}
  S.J. Linz,
  Phys. Rev. E {\bf 54}, 2925 (1996).

\bibitem{Edwards}
  S.F. Edwards and D.V. Grinev,
  Phys. Rev. E {\bf 58}, 4758 (1998).

\bibitem{Brey}
  J.J. Brey, A. Prados and B. S\'anchez-Rey,
  Phys. Rev. E {\bf 60}, 5685 (1999).

\bibitem{Gavrilov}
  K.L. Gavrilov,
  Phys. Rev. E {\bf 58}, 2107 (1998).

\bibitem{Philippe}
  P. Philippe and D. Bideau,
  Phys. Rev. E {\bf 63}, 051304 (2001).

\bibitem{Head1}
  D.A. Head and G.J. Rodgers,
  J. Phys. A {\bf 31}, 107 (1998).

\bibitem{Head2}
  D.A. Head,
  Phys. Rev. E {\bf 62}, 2439 (2000).

\bibitem{Luding}
  S. Luding, M. Nicolas and O. Pouliquen,
  in {\it Compaction of soils, Granulates and Powders}, D. Kolymbas, W. Fellin
  and A.A. Balkema (eds.), p. 241, Rotterdam (2000).

\bibitem{Hertzsch}
  J.-M. Hertzsch,
  Eur. Phys. J. B {\bf 18}, 459 (2000).

\bibitem{Nicodemi1}
  M. Nicodemi, A. Coniglio and H.J. Herrmann,
  Phys. Rev. E {\bf 55}, 3962 (1997).

\bibitem{Nicodemi2}
  M. Nicodemi,
  Phys. Rev. Lett. {\bf 82}, 3734 (1999).

\bibitem{Nicodemi3}
  M. Nicodemi,
  Physica A {\bf 285}, 267 (2000).

\bibitem{Caglioti}
  E. Caglioti, V. Loreto, H.J. Herrmann and M. Nicodemi,
  Phys. Rev. Lett. {\bf 79}, 1575 (1997).

\bibitem{Barrat1}
  A. Barrat and V. Loreto,
  J. Phys. A {\bf 33}, 4401 (2000).

\bibitem{Barrat2}
  A. Barrat, J. Kurchan, V. Loreto and M. Sellitto,
  Phys. Rev. Lett. {\bf 85}, 5034 (2000).

\bibitem{Sellitto}
  M. Sellitto and J.J. Arenzon,
  Phys. Rev. E {\bf 62}, 7793 (2000).

\bibitem{Kob}
  W. Kob  and H. C. Andersen, 
  Phys. Rev. E {\bf 48}, 4364 (1993).

\bibitem{Arenzon}
  Y. Levin , J.J. Arenzon and M. Sellitto, 
  Europhys. Lett. {\bf 55}, 767 (2001). 

\bibitem{Pouliquen}
  O. Pouliquen, M. Nicolas and P.D. Weidman,
  Phys. Rev. Lett. {\bf 79}, 3640 (1997).

\bibitem{Nicolas}
  M. Nicolas, P. Duru and  O. Pouliquen,
  Eur. Phys. J. E {\bf 3}, 309 (2000).

\bibitem{Villarruel}
  F.X. Villarruel, B.E. Lauderdale, D.M. Mueth and H.M. Jaeger, 
  Phys. Rev. E {\bf 61}, 6914 (2000).

\bibitem{Fusco1} 
  C. Fusco, P. Gallo, A. Petri and M. Rovere,
  Phys. Rev. E {\bf 65}, 026127 (2002).

\bibitem{Kasteleyn}
  P.W. Kasteleyn, Physica {\bf 27}, 1209 (1961).

\bibitem{Fisher}
  M.E. Fisher, Phys. Rev. {\bf 124}, 1664 (1961).

\bibitem{Fusco2} 
  C. Fusco,  P. Gallo, A. Petri and M. Rovere,
  J. Chem. Phys. {\bf 114}, 7563 (2001).

\bibitem{Duran}
  J. Duran, T. Mazozi, E. Clement and J. Rajchenbach,
  Phys. Rev. E {\bf 50}, 3092 (1994). 

\end{thebibliography}
\end{document}